- Considers the key features of the software package that has been developed and implemented. The package implement the I&R system's model, which has been discussed above. The distinguishing features of these tools are below.
    1. The package provides flexible and simple mechanism for support interaction between HTTP server and data, which have a tabular presentation.
    2. The package allows a cross-platform data transfer. The formats of templates of the text documents and the formats of the tabular data do not depend on the choice of the operation system. It is also possible to transfer the processor of the text document templates between various operation systems.
    3. The set of the resulting documents can be easily extended by mean ether changing the tabular data or extending the set of document templates.

Some intermediate project were already implemented using the conceptual model of the I&R system and the software package those have been discussed above. These projects were published as small and middle WWW sites in Internet and Intranet networks.

## Author's Information

**Andrii Donchenko** – Bonus Technology, Ph.D., Project Manager/Team Leader; 86D Bozhenko Str., Kyiv, Ukraine, e-mail: andriy.donchenko@gmail.com

# USING ELEMENTS OF SEMANTIC PARSING IN E-LEARNING ENVIRONMENTS

## Andrii Struik

*Abstract: Possibilities for using semantic parsing to estimate the correspondence of text materials to teaching aims, correspondence of test task to theoretical materials and other problems arising during the distance course designing and educational process itself in e-learning environments.*

*ACM Classification Keywords: K.3.1 Computer and Education: Computer uses in education*

## Introduction

A lot of researches are devoted to the possibilities of applying artificial intelligence tools in educational systems. The analysis shows that much attention is paid to semantic analysis. In particular there are many tools that use semantic analysis in automated assessment of student's complete answer to some question. Investigations conducted by the Modeling and Software Department of Kryvyi Rih Technical University show that possibilities of semantic analysis application in educational environments are considerably wider. In particular at the stage of distance courses design using semantic parsing it is possible to estimate the correspondence of text materials to teaching aims, correspondence of test tasks to theoretical materials used. During the process of training the elements of semantic parsing can be used for estimation of richness of messages content. The present article is devoted to the realization of these and other possibilities.

## Possibilities for Semantic Parsing Application in E-learning Environments

The tools of key information extraction from the text and analysis of its semantic context lay in the foundation of the methods proposed by us. A lot of investigations (among them there are many works by Peter Turney [Turney, 1997; Turney, 2000]) have been devoted to key information extractions tools from the text, compiling of short summaries and annotations on the basis of this information. The elements of such investigations are widely implemented in searching web-systems, electronic libraries catalogue tolls, but almost no investigations have been conducted for implementations of similar tools in the distance learning systems. At the same time information exchange in the interactive modules of educational environments takes place using text information and its automatic analysis could considerably increase the efficiency of educational process.

Several tasks that can by solved using elements of semantic parsing have been distinguished in our investigations. Here are some of them.

Firstly, the distinguishing of the key information is in itself an important tool and can be widely used in the distance learning organization. The present method enables, for example, to automate the process of keywords distinguishing in educational texts and glossary building, the creating of concise context of each chapter and of the whole educational course. And the most important is that the method of key information distinguishing in the educational text is a basis for all other directions of our investigations and before considering the other possibilities of semantic parsing application we concentrated our efforts on the development and perfection of this very module.

Secondly, key information distinguishing in the educational text allows creating the analyzer of educational material correspondence to the set educational aims. This will allow the developer of a distance course, firstly, to pay attention to the aims to be achieved in the distance course. Strictly determined aims enable the correspondence of educational materials to the set aims to be traced automatically, to indicate to the completeness of one or another question disclosure and to conduct search for the necessary information in the proposed database even in automatic mode. A collection of text messages, files or World Wide Web resources can be such a database. Thus the semantic parsing becomes an indispensable assistant during designing and creating the distance course in the educational environment.

Thirdly, the conditions for correspondence analysis of test tasks to the aims of education and to the contents of theoretical materials are being created. There are some well-known facts when much attention has been paid to insignificant aspects considered in the theoretical material during composing the test tasks. This, of course, decreased the efficiency of testing and did not give the objective estimation of students' knowledge. On the basis of the key information distinguishing it is possible to determine the correspondence of test tasks to the main contents of educational material and to show the developer of knowledge control system the possibility of error. Insufficient number of questions concerning the subject matter, having been distinguished by semantic parsing as a key one, can become another problem while compiling test tasks. The module of test tasks analysis will be able to indicate the non-correspondence between theoretical material and test tasks, and, thus, to contribute to the increase of testing efficiency.

Fourthly, there is an opportunity of semantic parsing application to estimate the correspondence of text messages that pass through interactive modules, to the subject of the current lesson. The important feature of distance courses is the intensive application of communication tools, such as e-mail, forum, chat to organize the communication between students and a teacher. Of course, such communication is of educational character and is used to estimate the quality of learning. That's why certain grades for activity in communication have been provided in the students' assessment system. But for such communication not only availability of text material but also its richness of contents and correspondence of the lesson subjects, that is being conducted at the moment, or the course subject as a whole, are important. The teacher has to estimate the richness of messages content by himself, that's why he/she faces the necessity to work through a great number of text messages daily. This is especially important during the intensive work with a great number of students. In this case the semantic parser allows automate the process of text messages analysis and determining their correspondence to the given theme or the course subject matter in general. This also creates the possibility to implement the automated knowledge control not only using testing but also using students' activity analysis in forum, e-mail or chat considering the richness of their messages content.

Semantic Parser Implementation and its Integration into E-learning Environments

As it has already been pointed, the procedure of key information distinguishing from the text messages, and in particular educational texts, underlies all the presented methods. Implementation of this procedure, proposed by us, is rather simple. First of all keywords are distinguished from the text on the basis of statistical data. Keywords determinations are performed on the basis of Zipf's laws. The implementation of them has become rather popular, for example, on modern search engines [Breslau, 1998]. Further, the context, where the keywords are used, is analyzed; a so called semantic cut for each keyword is prepared. A table from words that go with keywords is formed, semantic links between keywords and words-satellites are established. At this stage the analysis of synonymous and homonymous variations with application of corresponding dictionaries is provided. The generation of tables for each keyword is concluded with a cluster analysis, during which links between certain terms on the basis of their context are determined.

These tables are the foundation for realization of methods described above. Key notions and terms will be in the knots of the table. Their definitions can be obtained from the analysis of their context and, thus, a glossary for each educational chapter and for the whole course can be created. While determining the correspondence of educational materials to the course aims, the search of distinguished keywords is done in the formulations of aims, their context is analyzed, the semantic cut is built for each of them and is compared with the correspondent semantic cut. Coefficient of coincidence allows proving correspondence or non-correspondence of educational materials or separate chapters to the course aims. Similarly the correspondence of test tasks to theoretical materials is established and the analysis of content richness of text messages in chat, forum and e-mail is conducted.

After detailed elaboration and implementation of the semantic parsing methods for educational texts, our research was directed to the development of actions to integrate the semantic parser into the existing educational environments and developments of dialog modes between analyzer and user.

## Conclusion

The conducted investigations enable us to build software using semantic parsing elements for effective solving of numerous problems arising during distance courses design and educational process organization itself in distance learning environments. We hope that the conducted investigations will be the basis of a more complex project, i.e. the elaboration of expert-system which acts as a methodologist while designing, filling with necessary material and implementation of distance courses.

## Author's Information

**Andrii Striuk** – Kryvyi Rih Technical University; XXII Partz'yizd st., 11, room 219, Kryvyi Rih, 50027, Ukraine; e-mail: andrey_stryuk@mail.ru